\documentclass[pre,aps,amsmath,amssymb,amsfonts,floatfix,superscriptaddress,showpacs,twocolumn,footinbib,longbibliography]{revtex4-1}
\usepackage{graphicx}
\usepackage{amsmath,amsfonts}
\usepackage{xcolor}
\usepackage{color}
\usepackage{mathtools}
\usepackage{eufrak}
\usepackage{pgfplots}
\usepackage{soul}
\usepackage{multirow}
\usepackage{upgreek}

\newcommand{\up}{\uparrow}
\newcommand{\dn}{\downarrow}
\newcommand{\kv}{\ensuremath{\mathbf{k}}}
\newcommand{\qv}{\ensuremath{\mathbf{q}}}

\newcommand{\ch}{\ensuremath{\text{ch}}}
\newcommand{\sz}{\ensuremath{\text{sp}}}

\newcommand{\trip}{\ensuremath{\text{t}}}
\newcommand{\sing}{\ensuremath{\text{s}}}

\usepackage{tikz}

\usetikzlibrary{calc}
\usetikzlibrary{decorations.pathmorphing}
\usetikzlibrary{decorations.pathreplacing}
\usetikzlibrary{decorations.markings}
\usetikzlibrary{shapes.misc}
\usetikzlibrary{shapes}
\usetikzlibrary{positioning}
\usetikzlibrary{snakes}
\usetikzlibrary{arrows}
\usetikzlibrary{fadings}
\usetikzlibrary{calc,arrows}
\tikzstyle{decision} = [diamond, draw, fill=blue!20, text width=4.5em, text badly centered, node distance=3cm, inner sep=0pt]
\tikzstyle{block} = [rectangle, draw, fill=blue!20, text width=7em, text centered, rounded corners, minimum height=5em]
\tikzstyle{line} = [draw, -latex']
\tikzstyle{cloud} = [draw, ellipse,fill=red!20, node distance=3cm, minimum height=2em]
\tikzstyle{overbrace style}=[decorate,decoration={brace,raise=2mm,amplitude=3pt}]
\tikzstyle{overbrace text style}=[font=\footnotesize, above, pos=.5, yshift=3mm]
\tikzset{snake it/.style={decorate, decoration=snake}}
\usetikzlibrary{3d}
\usepackage{mathtools}
    \tikzset{
            partial ellipse/.style args={#1:#2:#3}{
                        insert path={+ (#1:#3) arc (#1:#2:#3)}
                            }
                        }

\usetikzlibrary{calc}
\tikzset{
            inertial frame/.style = {x={(-20:2cm)}, y={(-160:2cm)}, z={(90:2cm)}},
              local frame/.style = {shift={(local origin)}, x={(40:.7cm)}, y={(150:.7cm)}, z={(105:.7cm)}}
          }

    \tikzset{middlearrow/.style={
                decoration={markings,
                            mark= at position 0.65 with {\arrow{#1}} ,
                                    },
                                            postaction={decorate}
                                                }
                                                }
\tikzset{cross/.style={cross out, draw, 
         minimum size=2*(#1-\pgflinewidth), 
                  inner sep=0pt, outer sep=0pt}}

\def\presuper#1#2%
  {\mathop{}%
   \mathopen{\vphantom{#2}}^{#1}%
   \kern-0.5\scriptspace%
   #2}

\usepackage{hyperref}
\hypersetup{colorlinks=true,breaklinks,linkcolor=blue,urlcolor=blue,citecolor=blue}
\begin{document}

    \pgfmathdeclarefunction{gauss}{2}{%
          \pgfmathparse{1/(#2*sqrt(2*pi))*exp(-((x-#1)^2)/(2*#2^2))}%
          }
    \pgfmathdeclarefunction{mgauss}{2}{%
          \pgfmathparse{-1/(#2*sqrt(2*pi))*exp(-((x-#1)^2)/(2*#2^2))}%
          }
    \pgfmathdeclarefunction{lorentzian}{2}{%
        \pgfmathparse{1/(#2*pi)*((#2)^2)/((x-#1)^2+(#2)^2)}%
          }
    \pgfmathdeclarefunction{mlorentzian}{2}{%
        \pgfmathparse{-1/(#2*pi)*((#2)^2)/((x-#1)^2+(#2)^2)}%
          }

\author{Friedrich Krien}
\affiliation{Institute for Solid State Physics, TU Wien, 1040 Vienna, Austria}
\author{Anna Kauch}
\affiliation{Institute for Solid State Physics, TU Wien, 1040 Vienna, Austria}
\author{Karsten Held}
\affiliation{Institute for Solid State Physics, TU Wien, 1040 Vienna, Austria}

\title{Tiling with triangles: parquet and $GW\gamma$ methods unified}

\begin{abstract}
The parquet formalism and Hedin's $GW\gamma$ approach are unified into a
single theory of vertex corrections,
{corresponding to an exact reformulation of the parquet equations in terms of boson exchange.}
The method has no drawbacks compared to previous parquet solvers
but has the significant advantage that the vertex functions decay quickly
with frequencies and with respect to distances in real space.
These properties coincide with the respective separation of the length and energy scales
of the two-particle correlations into long/short-ranged and high/low-energetic.
\end{abstract}

\maketitle
  
\section{Introduction}
The systematic calculation of  vertex corrections in electronic systems historically builds upon two distinct formalisms,
the parquet formalism of De Dominicis and Martin~\cite{Dominicis64-2} (introduced also earlier by Diatlov et al. for meson scattering~\cite{Diatlov57})
and the $GW\gamma$ method introduced by Hedin~\cite{Hedin65}.
Although both approaches are in widespread use since the 1960's, they have largely remained separate entities.

The parquet approach~\cite{Bickers04,Tam13,Rohringer12,Valli15,Li16,Kauch19,Astretsov19}
classifies  vertex corrections into three scattering channels,
allowing an unbiased competition between the bosonic fluctuations in these channels~\cite{Friederich11,Metzner12,Tagliavini19,Kauch19-2, Pudleiner19-2}.
The Hedin equations, on the other hand, aim at the particle-hole channel, with vertex  corrections $\gamma$ in this channel being calculated self-consistently from the derivative of the self-energy with respect to the Green's function $\delta\Sigma/\delta G$~\cite{Onida02,Held11}.
Both formalisms constitute an exact quantum field theoretical framework,
but in practice one relies on approximations: In {the} case of the parquet formalism,  the
fully irreducible vertex $\Lambda$   is approximated, e.g., by $\Lambda=U$ (the bare interaction) in the parquet approximation~\cite{Bickers04} or by $\Lambda={\rm local}$ in the dynamical vertex approximation~\cite{Toschi07,Katanin09,Rohringer18}. In {the} case of the Hedin approach, $\gamma$ is  approximated, e.g.,  by $\gamma=1$ in the $GW$ approximation or by simple approximations in so-called  $GW\gamma$ approaches, even allowing for realistic materials calculations~\cite{Godby88,Aryasetiawan98,Onida02,Biermann03,Held11,Tomczak17,Nilsson17,Maggio17}.

One difference is that  the parquet formalism is formulated in terms of four-point electron-electron vertices (Feynman diagrammatic ``squares'', capitalized symbols in our notation), whereas Hedin~\cite{Hedin65} formulated his equations in terms of three-point electron-boson vertices (``triangles'', lowercase symbols). The latter can be reformulated easily in terms of four-point squares, see e.g.~\cite{Held11}, but to the best of our knowledge the parquet approach evaded hitherto a three-point (triangle) reformulation.

A second major difference between the two approaches is that the parquet equations
naturally obey the crossing symmetry but typical approximations violate conservation laws~\cite{Smith92,Janis99,Bickers04,Janis05,Janis17,Rohringer18,Kugler18,Krienthesis}, whereas  $GW\gamma$ approximations conversely often obey conservation laws~\cite{Almbladh99} but violate the crossing symmetry, and thereby the Pauli principle.
Indeed, only the exact solution satisfies the crossing symmetry and the Ward identity
at the same time, see Ref.~\cite{Kugler18} for a formal proof.

Aspects of both concepts come into play in the theory of collective bosonic fluctuations in fermionic systems, see, for example, Ref.~\cite{Chubukov14},
in particular of those in superconductors~\cite{Larkin08},
where the three-legged fermion-boson coupling and the screened interaction are used to construct four-point vertex corrections.
However, these considerations are almost always of a phenomenological kind and only a few 
Feynman diagrams of interest are calculated, such as the Aslamazov-Larkin
vertex correction~\cite{Aslamazov68,Bergeron11} (see diagram (b) in Fig.~\ref{fig:aslamazov} below).
But in terms of an overarching theory the relation between the parquet and Hedin formalisms remains, even after more than 50 years, only a tentative one.

In this paper, both approaches and viewpoints are merged into a single theory.
It is shown that the parquet decomposition of the vertex function,
which relies on the reducibility with respect to Green's functions~\cite{Rohringer12},
can be combined with the recently introduced {\sl single-boson exchange (SBE)} decomposition~\cite{Krien19-2}
that is based on the idea of reducibility with respect to the interaction, which generalizes the Hedin equations.
In particular, the diagrams that are reducible with respect to the interaction
can be removed exactly from the parquet expressions and suitable
ladder equations can be defined which replace the Bethe-Salpeter equations.
Through this {exact} reformulation {of the parquet method,} we tile our
diagrammatic ``floor'' not with conventional four-leg parquet ``squares'' but with three-leg ``triangles'',
with the exception being an irreducible ``square'' $\tilde{\Lambda}=\Lambda-U$, which vanishes in the parquet approximation.

The present paper is in close conjunction with Ref.~\cite{Krien20} where the parquet equations for  dual fermions~\cite{Rubtsov08} have been reformulated. Instead, here we show how the standard parquet approach~\cite{Bickers04,Tam13,Rohringer12,Valli15,Li16,Kauch19,Astretsov19} can be rewritten and connected with the Hedin equations.
{As a result, the self-energy of the parquet approach assumes the
``$GW\gamma$'' form, which is not the case for the parquet dual fermions~\cite{Krien20-2}.}
An efficient real fermion parquet solver for a quantum impurity model is presented and made available~\cite{KrienBEPSpython}.
Similar as  for dual fermions~\cite{Krien20},
this reformulation leads to a substantially improved feasibility of the parquet solution
because it removes simultaneously the high-frequency asymptotics~\cite{Wentzell20}
and the long-ranged fluctuations~\cite{Rohringer18} from the parquet equations.

The paper is organized as follows.
The Hedin and parquet formalisms are recollected in Sections~\ref{sec:hedin} and~\ref{sec:parquet}, respectively.
The two concepts are connected and merged in Section~\ref{sec:relation}.
A {unified} calculation scheme {using the SBE decomposition} is presented in Section~\ref{sec:unified}; and Section~\ref{sec:application} presents the implementation for a quantum impurity model {(Section~\ref{sec:AL})} {and examples for the
lattice Hubbard model from the parquet approximation} (Section~\ref{sec:PA}) using the \textit{victory} code~\cite{Kauch19}.
Further, in Section~\ref{sec:GW}, we reduce the results of the parquet approximaton step-by-step to the $GW$ approximation.
{We conclude in Sec.~\ref{sec:conclusions},
where we also discuss similarities and differences of the method
compared to the one presented in Ref.~\cite{Krien20}.}

\section{Hedin's formalism}\label{sec:hedin}
In Hedin's theory the self-energy of the electronic system is expressed in terms of
the Green's function $G$, the screened interaction $W$, and the vertex  $\gamma$.
For a single-band Hubbard-type system with the interaction $Un_\up n_\dn$
the self-energy can be expressed in the paramagnetic case as follows~\footnote{We use a Fierz splitting of $\frac{1}{2}$ between charge and spin channels~\cite{Krien19-3}.}:
\begin{align}
    \Sigma_k=\frac{U\langle n\rangle}{2}-\frac{1}{2}\sum_q G_{k+q}\left[W^\ch_q\gamma^\ch_{kq}+W^\sz_q\gamma^\sz_{kq}\right].
    \label{eq:hedin}
\end{align}
Here, $\ch$ and $\sz$ denote the charge and spin (or density and magnetic) combinations of the spin indices, respectively, see e.g.~\cite{Rohringer18};
$\langle n\rangle$ is the density; $k=(\kv,\nu)$ and $q=(\qv,\omega)$
denote fermionic and bosonic momentum-energy four-vectors, respectively,
$\nu,\omega$ are Matsubara frequencies.
Summations over $k,q$ imply a factor $T,\frac{1}{N}$ where
$T$ is the temperature and $N$ the number of lattice sites.
The {\sl Hedin vertex} $\gamma^{\ch (\sz)}$ takes,  in the exact theory,
 all vertex corrections  in the particle-hole channel into account.

The screened interaction $W$ corresponds to the bare Hubbard interaction
$U^\ch=U, U^\sz=-U, U^\sing=2U, U^\trip=0$ in the charge ($\ch$) and spin ($\sz$), singlet ($\sing$), triplet ($\trip$) channel, respectively, dressed by the  polarization $\Pi$, i.e.,
\begin{align}
W^{\ch/\sz}_q=\frac{U^{\ch/\sz}}{1-U^{\ch/\sz}\Pi^{\ch/\sz}_q},\;\;\;
W^\sing_q=\frac{U^\sing}{1-\frac{1}{2}U^\sing\Pi^\sing_q}\label{eq:w}.
\end{align}
For later use, we here introduced a $W^\sing$ also for the singlet particle-particle channel, while $W^\trip=0$. Both are not used in Hedin's original $GW\gamma$ approach, but are needed for the later connection to the parquet approach, which includes the particle-particle channel. The third, transversal particle-hole, channel is related to the particle-hole channel by crossing symmetry. Hence we do not need to introduce two further $W$'s and $\gamma$'s;   $W^{\ch (\sz)}$ and $\gamma^{\ch (\sz)}$ in the particle-hole channel are sufficient.

The polarization  in Eq.~(\ref{eq:w}) is in turn given by the Hedin vertex:
\begin{align}
\Pi^{\ch/\sz}_q=\sum_{k}G_k G_{k+q}\gamma^{\ch/\sz}_{kq},
\Pi^\sing_q=\sum_{k}G_k G_{q-k}\gamma^\sing_{kq}.\label{eq:pi}
\end{align}
Equations~\eqref{eq:hedin}-\eqref{eq:pi} are formally exact,
but in general the vertex corrections contained in $\gamma$ are unknown.
Hedin~\cite{Hedin65} suggested to calculate these through the Ward identity (see also Sec.~\ref{sec:unified}) but this is hardly feasible in practice.
Instead, the vertex corrections $\gamma$ are often neglected which gives rise to the eponymous $GW$ approximation.
If (some approximate) vertex corrections are kept one speaks of a $GW\gamma$ approach~\cite{Sole94}.

The diagrammatic background to introduce $W$ and $\Pi$ in the Hedin equations is the concept of interaction-(ir)reducibility:
a Feynman diagram is interaction-reducible if and only if it separates into two pieces if one interaction line is cut out.
Eventually, we need to consider all  vertex corrections, i.e., the full vertex function $F^\alpha_{kk'q}$.
The interaction-reducible diagrams of $F$ take the form~\cite{Krien19-2},
\begin{align}
    \Delta^\alpha_{kk'q}=\gamma^\alpha_{kq}W^\alpha_q\gamma^\alpha_{k'q}.\label{eq:nabla}
\end{align}
Quite obviously, we can cut an interaction line $U$ within $W$ and hence $\Delta$, and vice versa any interaction-reducible diagram in channel $\alpha$ has to be of the form Eq.~(\ref{eq:nabla}).
The vertex $\Delta$ has been coined {\sl single-boson exchange} (SBE) vertex \cite{Krien19-2} as it involves the exchange of a single boson with four-vector $q$ within $W$.

These interaction-reducible contributions must not be contained in $\gamma$, and hence  must be subtracted from $F$ to avoid a double counting~\footnote{\label{foot:ladder}
Due to this {\sl interaction-irreducible} property of the Hedin vertex
it satisfies the ladder equations~\cite{Hedin65},
\begin{align*}
\gamma^{\ch/\sz}_{kq}=&1+\sum_{k'}S^{\ch/\sz}_{kk'q}G_{k'}G_{k'+q}\gamma^{\ch/\sz}_{k'q},\\
\gamma^\sing_{kq}=&-1-\frac{1}{2}\sum_{k'}S^\sing_{kk'q}G_{k'}G_{q-k'}\gamma^\sing_{k'q},
\end{align*}
where $S$ is the corresponding Bethe-Salpeter kernel {\sl without}
the bare interaction, defined in Eq.~\eqref{eq:kernel_u}.}. This yields~\cite{Krien19-2}:
\begin{subequations}
\begin{align}
\gamma^{\ch/\sz}_{kq}=&1+\sum_{k'}(F^{\ch/\sz}_{kk'q}-\Delta^{{\ch/\sz}}_{kk'q})
G_{k'}G_{k'+q},\label{eq:gamma_chsp}\\
\gamma^\sing_{kq}=&-1+\frac{1}{2}\sum_{k'}(F^\sing_{kk'q}-\Delta^{\sing}_{kk'q})
G_{k'}G_{q-k'}\label{eq:gamma_singlet}.
\end{align}
\label{eq:gamma}
\end{subequations}
Here, the Green's functions serve the conversion of the four point vertex $F-\Delta$
to the three point vertex $\gamma$, and the ``1'' generates in the Hedin formulations the
contributions without vertex corrections.
{There is no triplet Hedin vertex because the bare interaction vanishes
in this channel, $U^\trip=0$.}

\section{Parquet formalism}\label{sec:parquet}
\begin{figure}
 \includegraphics[width=0.45\textwidth]{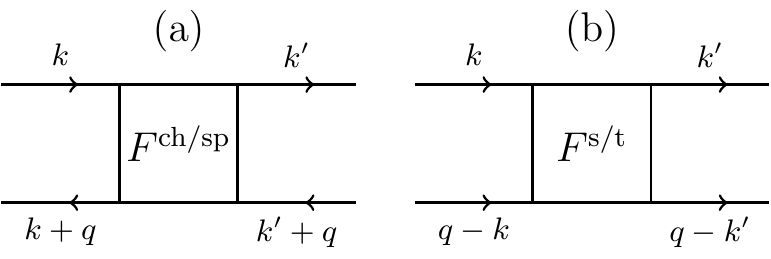}
    \caption{\label{fig:4vertex} Label convention for (a) the particle-hole and
    (b) the particle-particle notation.}
    \end{figure}

The parquet formalism~\cite{Dominicis64,Dominicis64-2,Bickers04,Gunnarsson16,Rohringer18} is based on the insight that the full vertex $F$ can be decomposed into the fully-irreducible vertex $\Lambda$ and reducible vertices $\Phi^r$ in the particle-hole
($r=ph$), transversal-particle-hole ($r=\overline{ph}$) and particle-particle-channel ($r=pp$).
Now (two-particle) irreducibility is to be understood with respect to  cutting two Green's function lines. Each Feynman diagram for $F$ belongs to exactly one of these four classes, i.e.,~$F=\Lambda+ {\Phi}^{ph}+{\Phi}^{\overline{ph}}+{\Phi}^{pp}$ \cite{Bickers04,Rohringer18}. In terms of the spin combinations $\alpha=\ch,\sz$,  we get with the momentum-convention for the particle-hole channel
(cf. Fig.~\ref{fig:4vertex}, left),
\begin{align}
    F^\alpha_{kk'q}=&{\Lambda}^{\alpha}_{kk'q}
    +{\Phi}^{ph,\alpha}_{kk'q}\label{eq:parquet}\\
    -&\frac{1}{2}{\Phi}^{ph,\ch}_{k,k+q,k'-k}
    -\frac{3-4\delta_{\alpha,\sz}}{2}{\Phi}^{ph,\sz}_{k,k+q,k'-k}\notag\\
    +&\frac{1-2\delta_{\alpha,\sz}}{2}{\Phi}^{pp,\sing}_{kk',k+k'+q}
    +\frac{3-2\delta_{\alpha,\sz}}{2}{\Phi}^{pp,\trip}_{kk',k+k'+q}.\notag
\end{align}
Here, we have expressed  $\Phi^{\overline{ph}}$ in terms of $\Phi^{{ph}}$ in the second line using the crossing relation~\cite{Rohringer12}, and properly translated the $\sing$ and $\trip$ components and momenta of the $pp$ channel in the third line.
The fully irreducible vertex ${\Lambda}$ or an approximation thereof,
such as the  {\sl parquet approximation} ${\Lambda}^{\alpha}=U^\alpha$,
serves as an input.

Since $\alpha=\ch,\sz$ and $\alpha=\sing,\trip$ already uniquely determine the
channel $r=ph$ and $r=pp$, respectively, we drop the channel index $r$ in the following.

There is only one $F$ with two independent spin combinations, but one can use
the singlet and triplet combinations and $pp$ momentum convention (cf. Fig.~\ref{fig:4vertex}),
which is related to the above by
\begin{subequations}
\begin{align}
    F^\sing_{kk'q}=&\frac{1}{2}\left(F^\ch_{kk',q-k-k'}-3F^\sz_{kk',q-k-k'}\right),\label{eq:singlet_chsp}\\
    F^\trip_{kk'q}=&\frac{1}{2}\left(F^\ch_{kk',q-k-k'}+F^\sz_{kk',q-k-k'}\right).\label{eq:triplet_chsp}
\end{align}
\end{subequations}
One can further introduce an irreducible vertex in the respective channel
\begin{equation}
  {\Gamma}^{\alpha}_{kk'q} = F^\alpha_{kk'q}-{\Phi}^{\alpha}_{kk'q}.~\label{eq:Gir}
\end{equation}

For calculating the reducible vertices, we employ the Bethe-Salpeter equations which  in terms of $\Phi$ read 
\begin{subequations}
\begin{align}
{\Phi}^{\ch/\sz}_{kk'q}
=&\sum_{k''}{\Gamma}^{\ch/\sz}_{kk''q}G_{k''}G_{k''+q}F^{\ch/\sz}_{k''k'q},
\label{eq:phi_ph}\\
{\Phi}^{\sing/\trip}_{kk'q}
=&\mp\frac{1}{2}\sum_{k''}{\Gamma}^{\sing/\trip}_{kk''q}
G_{k''}G_{q-k''}F^{\sing/\trip}_{k''k'q}.\label{eq:phi_pp}
\end{align}
\label{eq:phi}
\end{subequations}
Here, we can replace ${\Gamma}$ by Eq.~(\ref{eq:Gir}), which allows for a self-consistent calculation of $\Phi$ and $F$ in the four-channels if $\Lambda$ is known as an input. Further, $G$ and $\Sigma$ can be calculated self-consistently as well,
using additionally the Dyson equation and Schwinger-Dyson equation
[that is equivalent to Eq.~\eqref{eq:hedin}].

\section{A unified approach to vertex corrections}\label{sec:relation}
We now relate the Hedin and parquet formalisms described in Sections~\ref{sec:hedin} and~\ref{sec:parquet}. Starting point is an analog to the parquet Eq.~(\ref{eq:parquet}) but formulated in terms of interaction-(ir)reducible vertices instead of the two-particle
(ir)reducibility of Eq.~\eqref{eq:parquet}.  
This SBE decomposition~\cite{Krien19-2} into
interaction-reducible channels reads for $\alpha=\ch,\sz$:
\begin{align}
    F^\alpha_{kk'q}=&{\Lambda}^{\text{Uirr},\alpha}_{kk'q}
    +{\Delta}^{\alpha}_{kk'q}\label{eq:jib_aim}\\
    -&\frac{1}{2}{\Delta}^{\ch}_{k,k+q,k'-k}
    -\frac{3-4\delta_{\alpha,\sz}}{2}{\Delta}^{\sz}_{k,k+q,k'-k}\notag\\
    +&\frac{1-2\delta_{\alpha,\sz}}{2}{\Delta}^{\sing}_{kk',k+k'+q}-2U^\alpha.\notag
\end{align}
The essential difference to the parquet Eq.~(\ref{eq:parquet}) is that  the vertices $\Delta^\alpha$ defined in Eq.~\eqref{eq:nabla} are reducible with respect
to the bare interaction $U^\alpha$~\footnote{
One may also say that the vertices $\Delta=\gamma W\gamma$ are reducible
with respect to the screened interaction $W$, however,
it is useful to consider also the bare interaction diagram
$U$ as {\sl reducible}~\cite{Krien19-2}.
Interestingly, the bare interaction does not play a role after we
have passed over to the parquet expression~Eq.~\eqref{eq:uirr_parquet},
which implies that $W$-reducibility is a meaningful concept {\sl after} the bosonization.
Indeed, in a functional formulation of the $GW$ approach the screened interaction
plays the role of a fundamental variable~\cite{Almbladh99}.}.
The bare interaction is itself interaction-reducible and hence included in the $\Delta^\alpha$'s; thus we need to subtract $2 U^\alpha$ in Eq.~\eqref{eq:jib_aim} to prevent an overcounting.
As already discussed in  Section~\ref{sec:hedin}, $U^\trip=W^\trip=\Delta^\trip=0$.
  
This also implies  $\Lambda^\text{Uirr}$ is fully irreducible with respect to the interaction, and  must \textit{not} be confused with the vertex $\Lambda$
of the parquet decomposition~Eq.~\eqref{eq:parquet}
which is fully irreducible with respect to pairs of Green's functions. This implies on the one hand that $U$ is contained in  $\Lambda$
but not in $\Lambda^\text{Uirr}$. But otherwise $\Lambda$ contains fewer diagrams than $\Lambda^\text{Uirr}$ as each diagram that is interaction reducible is also two-particle  reducible since we can cut the two Green's functions on one side of the two-particle interaction instead of the interaction itself~\footnote{Compare also Figs. 3 and 4 in Ref.~\cite{Krien19-2}.
In other words, interaction reducibility {\sl implies} two-particle reducibility,
with the only exception of the bare interaction itself, which is (fully) two-particle irreducible.}.

In the following we will relate the parquet equation~\eqref{eq:parquet} and the SBE generalization  Eq.~(\ref{eq:jib_aim}) of the Hedin formalism, and formulate a unified theory. To this end, we will pinpoint the difference between $\Lambda^\text{Uirr}$ and $\Lambda$, which is denoted as~\cite{Krien20}  {\em multi-boson exchange} (MBE) diagrams $M^\alpha$ (the SBE diagrams $\Delta^\alpha$ are not part of  $\Lambda^\text{Uirr}$; Fig.~\ref{fig:aslamazov} below clarifies the multi-boson character of $M$). We will derive the equations to calculate $M^\alpha$ and  $\Delta^\alpha$ self-consistently in a unified Hedin and parquet formalism. This approach, while fully equivalent to the parquet approach, is formulated with the Hedin vertices and screened interactions and bears the advantage that the calculated vertex functions $\Lambda^\text{Uirr}, M$ decay with the frequencies and depend only weakly on the momenta, compared to $F, \Phi$ of the original parquet approach.

First, we start with some definitions. Analogously to $\Gamma^{\alpha}$ in Eq.~\eqref{eq:Gir},
we introduce  vertices $T^{\alpha}$ that are irreducible with respect
to the bare interaction $U^\alpha$ only in a particle-hole channel ($\alpha=\ch,\sz$ ) or
in a particle-particle channel ($\alpha=\sing,\trip$)
by removing the reducible diagrams ${\Delta}^{\alpha}$ in that channel:
\begin{align}
{T}^{\alpha}_{kk'q} =&F^\alpha_{kk'q} - {\Delta}^{\alpha}_{kk'q}.\label{eq:sbe_bse}
\end{align}
By comparison with Eqs.~\eqref{eq:gamma_chsp} and~\eqref{eq:gamma_singlet} we see that
the vertices $T$ describe the vertex corrections for the Hedin vertex $\gamma$.
The latter is therefore also irreducible with respect
to the bare interaction in the corresponding channel~\cite{Krien19-2,Rohringer16}.

As is the custom in Hedin's formalism we remove the bare interaction $U^\alpha$ from the irreducible vertex $\Gamma^{\alpha}$:
\begin{align}
S^{\alpha}=\Gamma^{\alpha}-U^\alpha.~\label{eq:kernel_u}
\end{align}
Now we collect all diagrams that are interaction-irreducible (but two-particle reducible) as the difference
\begin{align}
{M}^{\alpha}_{kk'q}=T^{\alpha}_{kk'q}-S^{\alpha}_{kk'q}.~\label{eq:bse_t}
\end{align}

\begin{figure}
\includegraphics[width=0.48\textwidth]{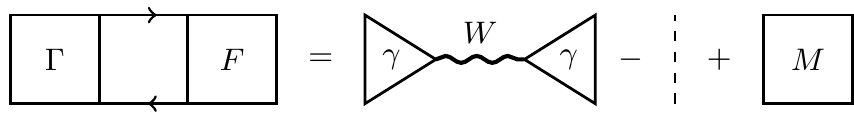}
\caption{\label{fig:phi}
Relation Eq.~\eqref{eq:phi_m} between the particle-hole reducible vertices
$\Phi$ in Eq.~\eqref{eq:phi_ph} and $M$ defined in Eq.~\eqref{eq:mbe_ph},
which represents multi-boson exchange (cf. Fig.~\ref{fig:aslamazov}).
Arrows and a dashed line denote Green's function $G$ and the bare interaction $U$, respectively.
}
\end{figure}

\begin{figure}
\begin{center}
\includegraphics[width=0.4\textwidth]{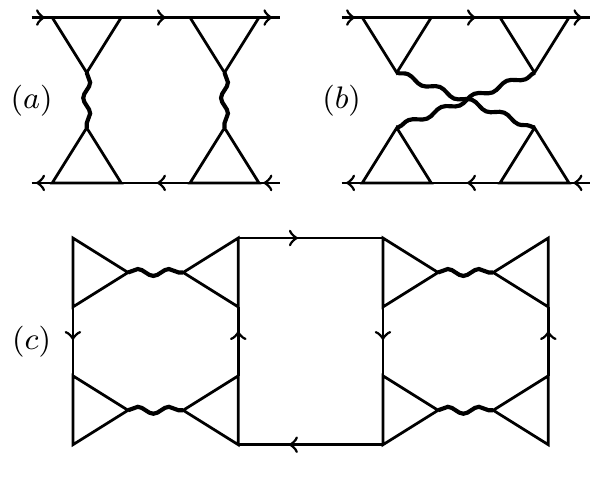}
\end{center}
    \caption{\label{fig:aslamazov}
    Tiling with triangles: exemplary vertex corrections corresponding to multiple boson exchange, see also Ref.~\cite{Krien20}.
    Diagrams (a) and (b) represent boson exchange in the particle-hole (a)
    and particle-particle (b) channels.
    Diagram (c) corresponds to a mixing of horizontal and vertical particle-hole channels.
    }
\end{figure}

Conversely, this means that $\Phi^{\alpha}$ consists of  ${M}^{\alpha}$ plus the interaction-reducible vertices $\Delta^{\alpha}$ in the respective channel,
\begin{align}
\Phi^{\alpha}_{kk'q}= F^{\alpha}_{kk'q}- \Gamma^{\alpha}_{kk'q} = \Delta^{\alpha}_{kk'q}-U^\alpha+{M}^{\alpha}_{kk'q}.\label{eq:phi_m}
\end{align}
Here again $U^\alpha$ needs to be subtracted as it is included
in $\Delta^{\alpha}$ but not in $\Phi^{\alpha}$.
A diagrammatic representation of Eq.~\eqref{eq:phi_m} is shown in Fig.~\ref{fig:phi}
for the particle-hole channel. Note that  $\Phi^{\trip}=M^{\trip}$ since $U^\trip=\Delta^{\trip}=0$.

With these definitions, we can now relate $\Lambda^{\text{Uirr},\alpha}$ of the SBE decomposition~Eq.~\eqref{eq:jib_aim} to ${\Lambda}^{\alpha}$ of the parquet~Eq.~\eqref{eq:parquet}, or more specifically to
\begin{align}
\tilde{\Lambda}^{\alpha}={\Lambda}^{\alpha}-U^\alpha.\label{eq:lambdatilde}
\end{align}
To this end, we equate Eq.~\eqref{eq:jib_aim} to Eq.~\eqref{eq:parquet}, which both yield $F^{\alpha}$, and express $\Phi^{\alpha}$ by $M^{\alpha}$ using~Eq.~\eqref{eq:phi_m}.
We are left with
\begin{align}
    \Lambda^{\text{Uirr},\alpha}_{kk'q}=&\tilde{\Lambda}^{\alpha}_{kk'q}
    +{M}^{\alpha}_{kk'q}\label{eq:uirr_parquet}\\
    -&\frac{1}{2}{M}^{\ch}_{k,k+q,k'-k}
    -\frac{3-4\delta_{\alpha,\sz}}{2}{M}^{\sz}_{k,k+q,k'-k}\notag\\
    +&\frac{1-2\delta_{\alpha,\sz}}{2}{M}^{\sing}_{kk',k+k'+q}
    +\frac{3-2\delta_{\alpha,\sz}}{2}{M}^{\trip}_{kk',k+k'+q}.\notag
\end{align}
All $\Delta^{\alpha}$'s cancel, as it must be.

We still need to calculate the ${M}^{\alpha}$'s. This can be done through Bethe-Salpeter-like equations similar as the ${\Phi}^{\alpha}$'s in Eq.~\eqref{eq:phi} of the original parquet formalism. 
Starting with Eq.~\eqref{eq:phi}, substituting $F^{\alpha}$,  $\Gamma^{\alpha}$ and $\Phi^{\alpha}$ by Eqs.~\eqref{eq:sbe_bse}, \eqref{eq:kernel_u}, and \eqref{eq:phi_m}, respectively,
and removing all interaction-reducible contributions from the left and right hand side, this yields
\begin{subequations}
\begin{align}
{M}^{\ch/\sz}_{kk'q}
=&\sum_{k''}{S}^{\ch/\sz}_{kk''q}G_{k''}G_{k''+q}T^{\ch/\sz}_{k''k'q},
\label{eq:mbe_ph}\\
{M}^{\sing/\trip}_{kk'q}
=&\mp\frac{1}{2}\sum_{k''}{S}^{\sing/\trip}_{kk''q}
G_{k''}G_{q-k''}T^{\sing/\trip}_{k''k'q}.\label{eq:mbe_pp}
\end{align}
\end{subequations}
Here, $T=S+M$ [Eq.\eqref{eq:bse_t}] can be substituted.
  
Besides this Bethe-Salpeter equation, we need the eponymous parquet equation, i.e., Eq.~\eqref{eq:parquet} in the original parquet formalism. Moving $\Phi^{\alpha}$ for the considered four channels ($\alpha$) to the left hand side in Eq.~\eqref{eq:parquet} and reexpressing everything in terms of the new variables ($\Lambda^\text{Uirr}, M, \Delta$),  we obtain, analogous to Ref.~\cite{Krien20}, the parquet equation formulated in terms of
\begin{widetext}
\begin{subequations}
\begin{align}
{S}^{\ch}_{kk'q}=\,&{\Lambda}^{\text{Uirr},\ch}_{kk'q}-{M}^{\ch}_{kk'q}
-\frac{1}{2}{\Delta}^{\ch}_{k,k+q,k'-k}-\frac{3}{2}{\Delta}^{\sz}_{k,k+q,k'-k}
+\frac{1}{2}{\Delta}^{\sing}_{kk',k+k'+q}-2U^\ch,\label{eq:ph_kernel_ch}\\
{S}^{\sz}_{kk'q}=\,&{\Lambda}^{\text{Uirr},\sz}_{kk'q}-{M}^{\sz}_{kk'q}
-\frac{1}{2}{\Delta}^{\ch}_{k,k+q,k'-k}+\frac{1}{2}{\Delta}^{\sz}_{k,k+q,k'-k}
-\frac{1}{2}{\Delta}^{\sing}_{kk',k+k'+q}-2U^\sz,\label{eq:ph_kernel_sp}\\
{S}^{\sing}_{kk'q}=\,&{\Lambda}^{\text{Uirr},\sing}_{kk'q}\;\,-{M}^{\sing}_{kk'q}
+\frac{1}{2}{\Delta}^{\ch}_{kk',q-k'-k}-\frac{3}{2}{\Delta}^{\sz}_{kk',q-k'-k}
+\frac{1}{2}{\Delta}^{\ch}_{k,q-k',k'-k}-\frac{3}{2}{\Delta}^{\sz}_{k,q-k',k'-k}-U^\ch+3U^\sz,\label{eq:pp_kernel_sing}\\
{S}^{\trip}_{kk'q}=\,&{\Lambda}^{\text{Uirr},\trip}_{kk'q}\;\,-{M}^{\trip}_{kk'q}
+\frac{1}{2}{\Delta}^{\ch}_{kk',q-k'-k}+\frac{1}{2}{\Delta}^{\sz}_{kk',q-k'-k}
-\frac{1}{2}{\Delta}^{\ch}_{k,q-k',k'-k}-\frac{1}{2}{\Delta}^{\sz}_{k,q-k',k'-k},\label{eq:pp_kernel_trip}
\end{align}
\end{subequations}
\end{widetext}
which we need as input for the Bethe-Salpeter-like Eqs.~\eqref{eq:mbe_ph} and \eqref{eq:mbe_pp}.

The expressions for the ladder kernels $S$ defined in
Eqs.~\eqref{eq:ph_kernel_ch}-\eqref{eq:pp_kernel_trip}
elucidate the physical picture implied in the reformulated parquet equations:
The parquet diagrams are reexpressed in terms of single- and multi-boson exchange,
where the latter is represented by $M$ which arises from the ladder Eq.~\eqref{eq:bse_t}
via repeated exchange of bosons, starting from the second order.
The feedback of $M$ on the ladder kernel $S$ leads to the channel
mixing that is characteristic of the parquet approach.
Feynman diagrams corresponding to multi-boson exchange are shown in Fig.~\ref{fig:aslamazov}.

{Let us emphasize that our unification of the parquet and $GW\gamma$ methods is a middle-ground   reformulation of both (exact) approaches. It is not a merger that combines elements of two approaches in a distinctively new method such as, e.g., $GW$+DMFT~\cite{Biermann03,Sun02}.
More closely related than $GW$+DMFT is the multi-loop flow equation
\cite{Kugler18} which extends the functional renormalization group
(fRG,~\cite{Metzner12}) to the parquet approach.}

\section{Calculation scheme}\label{sec:unified}
Now we are in a position to formulate the BEPS calculation scheme, which was introduced for dual fermions in Ref.~\cite{Krien20}.
The algorithm is as follows (for clarity, we repeat the most relevant equations):

\noindent{\textbf{Step 0} (starting point):}
Choose an approximation for $\tilde{\Lambda}$
(parquet approximation: $\tilde{\Lambda}\equiv0$; D$\Gamma$A: $\tilde{\Lambda}={\rm local}$).
Make an initial guess for the self-energy $\Sigma$,
polarization $\Pi$, Hedin vertices $\gamma$, and the MBE vertices $M$.
\vspace{.2cm}

\noindent{\textbf{Step 1}:}
Update the propagators (Green's function and screened interaction)
\begin{align}
    G_k=&\frac{G^0_k}{1-G^0_k\Sigma_k},\label{eq:cc_dyson}
\end{align}
\begin{subequations}
\begin{align}
    W^{\ch/\sz}_q=&\frac{U^{\ch/\sz}}{1-U^{\ch/\sz}\Pi^{\ch/\sz}_q}\label{eq:cc_dyson_w},\\
    W^\sing_q=&\frac{U^\sing}{1-\frac{1}{2}U^\sing\Pi^\sing_q},
\end{align}
\end{subequations}
where $G^0$ is the non-interacting Green's function.
\vspace{.2cm}

\noindent{\textbf{Step 2:}}
Obtain the interaction-reducible vertex 
\begin{align}
    \Delta^\alpha_{kk'q}=\gamma^\alpha_{kq}W^\alpha_q\gamma^\alpha_{k'q}.\label{eq:nabla_2}
\end{align}
\vspace{.2cm}

\noindent{\textbf{Step 3:}}
Calculate the  irreducible  kernel $S$ from Eqs.~\eqref{eq:ph_kernel_ch}-\eqref{eq:pp_kernel_trip}, where   $\Lambda^\text{Uirr}$  is obtained from $M$ and the fixed $\tilde{\Lambda}$ through~Eq.~\eqref{eq:uirr_parquet}.
\vspace{.2cm}

\noindent{\textbf{Step 4}:}
With this $S$  solve the ladder equations
\begin{subequations}
\begin{align}
{M}^{\ch/\sz}_{kk'q}
=&\sum_{k''}{S}^{\ch/\sz}_{kk''q}G_{k''}G_{k''+q}T^{\ch/\sz}_{k''k'q},
\label{eq:mbe_ph_2}\\
{M}^{\sing/\trip}_{kk'q}
=&\mp\frac{1}{2}\sum_{k''}{S}^{\sing/\trip}_{kk''q}
G_{k''}G_{q-k''}T^{\sing/\trip}_{k''k'q},\label{eq:mbe_pp_2}
\end{align}
\end{subequations}
using $T^{\alpha}_{kk'q}=S^{\alpha}_{kk'q}+{M}^{\alpha}_{kk'q}$.
\vspace{.2cm}

\noindent{\textbf{Step 5}:}
Update the Hedin vertices
\begin{subequations}
\begin{align}
\gamma^{\ch/\sz}_{kq}=&1+\sum_{k'}(F^{\ch/\sz}_{kk'q}-\Delta^{{\ch/\sz}}_{kk'q})G_{k'}G_{k'+q},
\label{eq:cc_gamma_chsp}\\
\gamma^\sing_{kq}=&-1+\frac{1}{2}\sum_{k'}(F^\sing_{kk'q}-\Delta^{\sing}_{kk'q})G_{k'}G_{q-k'}.
\label{eq:cc_gamma_singlet}
\end{align}
\end{subequations}
Here, $F$ is expressed through the SBE decomposition~Eq.~\eqref{eq:jib_aim}
and the parquet expression~Eq.~\eqref{eq:uirr_parquet}.
\vspace{.2cm}

\noindent{\textbf{Step 6}:}
Update the self-energy and polarization
\begin{align}
    \Sigma_k=\frac{U\langle n\rangle}{2}-\frac{1}{2}\sum_q G_{k+q}\left[W^\ch_q\gamma^\ch_{kq}+W^\sz_q\gamma^\sz_{kq}\right],\label{eq:cc_hedin}
    \end{align}

\begin{subequations}
\begin{align}
\Pi^{\ch/\sz}_q=&\sum_{k}G_k G_{k+q}\gamma^{\ch/\sz}_{kq},\\
\Pi^\sing_q=&\sum_{k}G_k G_{q-k}\gamma^\sing_{kq}.\label{eq:cc_polarization}
\end{align}
\end{subequations}
Iterate steps 1 to 6 until convergence.
\vspace{.2cm}

In Step 3 the ladder kernel $S$ is calculated {\sl on-the-fly}
for only one bosonic momentum-energy $q$ at a time.
In Step 4 the vertices $T$ need not be evaluated, only $M$ are stored.
As a result, the Hedin vertices in Eqs.~\eqref{eq:cc_gamma_chsp} and~\eqref{eq:cc_gamma_singlet} can be expressed in terms of $\tilde{\Lambda}, \Delta,$ and $M$.
Only the quantities mentioned in Step 0 need to be stored and updated over the iterations.

\begin{figure*}
  \includegraphics[width=0.95\textwidth]{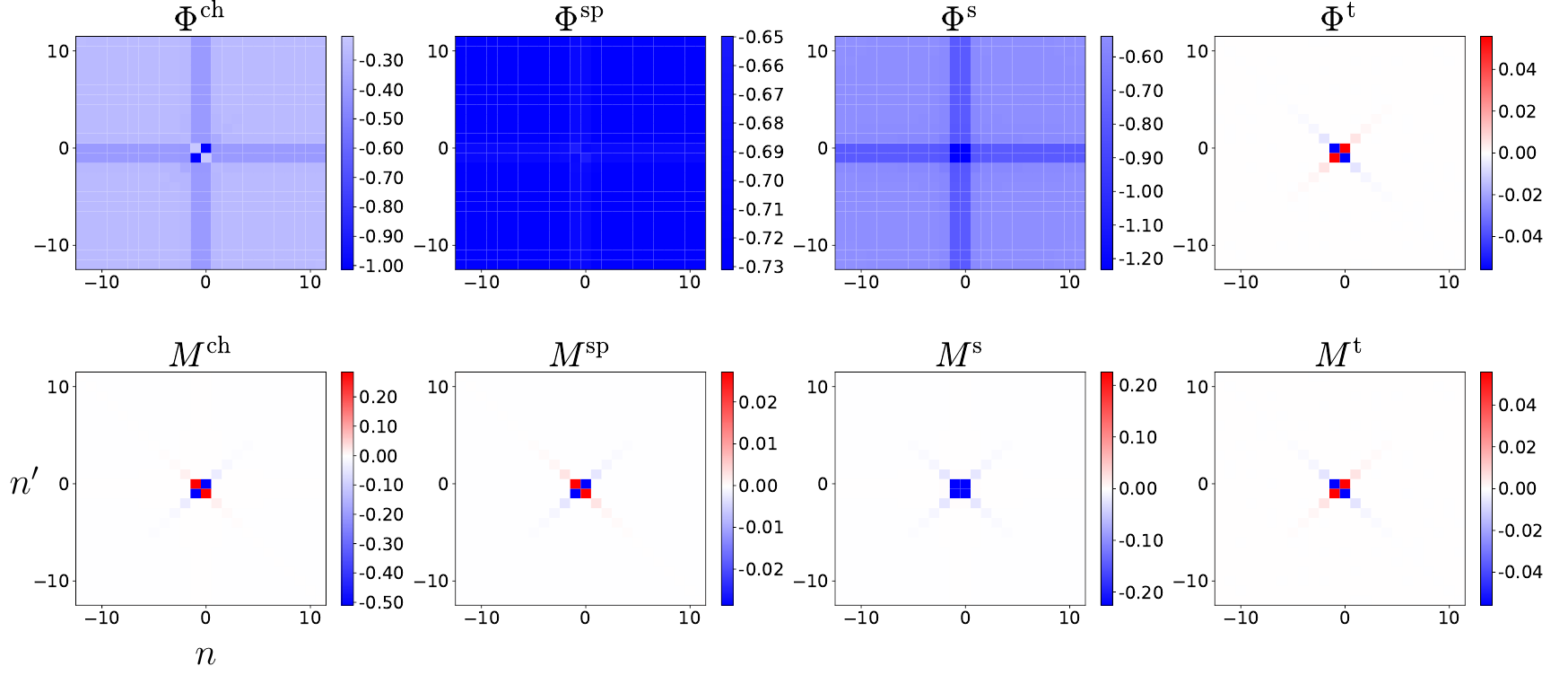}
  \caption{\label{fig:vertices}
  Reducible vertex functions of the atomic limit at $U/T=2$.
  Axes show the fermionic Matsubara indices ($\omega=0$).
  Top: $\Phi$ corresponding to the original parquet decomposition~Eq.~\eqref{eq:parquet}.
  Bottom: Vertices $M$ of the parquet expression~Eq.~\eqref{eq:uirr_parquet}.
  }
\end{figure*}

\paragraph*{\textbf{Relation to Hedin's equations:}}
The calculation scheme above differs from Hedin's original work
through the prescription for the ladder kernel $S$ in Step 3.
In Hedin's equations~\cite{Onida02,Held11} the ladder kernel is given
by the functional derivative $S=\delta(\Sigma-\Sigma^H)/\delta G$
where $\Sigma^H=U\langle n\rangle/2$.
With this $S$ and using $F-\Delta=T$ in Eq.~\eqref{eq:cc_gamma_chsp} the algorithm
is equivalent to Hedin's equations.
This functional derivative is however difficult to calculate in practice.
Here, instead, $S$ is obtained from the parquet diagrams in Step 3, as proved in Sec.~\ref{sec:relation}.

{
\section{{Numerical examples}}\label{sec:application}
In this section we evaluate the key quantities 
that play a role in the efficient calculation scheme defined in Sec.~\ref{sec:unified} (e.g., $W,\gamma,M$)
and demonstrate the low-energetic and short-ranged properties of the corresponding vertices $M$.
As concrete examples we consider the exact solution of the atomic limit and the parquet approximation
for the lattice Hubbard model at weak coupling. 

Here, the results for the atomic limit have been obtained using the corresponding implementation made available with this paper~\cite{KrienBEPSpython}; the relevant quantities for the Hubbard model have been evaluated using the \textit{victory} implementation of the traditional parquet equations~\cite{Kauch19}. 
}

\subsection{Atomic Limit}
\label{sec:AL}
We apply the BEPS method to a toy model, the atomic limit of e.g.\ the Hubbard model at half-filling.
This model is exactly solvable and it has a nontrivial solution for the vertex functions.
Analytical expressions for all components of the
parquet decomposition~Eq.~\eqref{eq:parquet} are available~\cite{Thunstroem18}.
Starting from the exact fully irreducible vertex $\Lambda_{\nu\nu'\omega}$
the calculation cycle
in Sec.~\ref{sec:unified} recovers the correlation functions of the atomic limit.

A \texttt{Python} implementation~\cite{KrienBEPSpython} is provided which converges
on a single core within a few minutes~\footnote{Due to the exponential difference between charge and spin fluctuations in the atomic limit the linear mixing used in the provided script
does not work for large values of $U/T$. A nonlinear root-finder improves the convergence~\cite{Krien19-3}.}.

We focus here on one advantage of the BEPS calculation scheme,
evident already in the atomic limit,
which is the decay of the vertex functions at high frequencies.
The top panels of Fig.~\ref{fig:vertices} show the reducible vertices
$\Phi_{\nu\nu'\omega}$ of the parquet decomposition in Eq.~\eqref{eq:parquet}.
The vertices belonging to the channels $\alpha=\ch,\sz,\sing$
have features that do not decay at high frequencies,
whereas the triplet vertex $\alpha=\trip$ decays.
This is the case because the bare interaction vanishes in the triplet channel, $U^\trip=0$.
The bottom panels of Fig.~\ref{fig:vertices} show the corresponding vertices $M_{\nu\nu'\omega}$
of the parquet expression~\eqref{eq:uirr_parquet}.
Evidently, all features of these vertices decay at high frequency (in {the} case of the triplet channel trivially because $M^{\trip}=\Phi^{\trip}$).

\subsection{Parquet approximation}
\label{sec:PA}
Next, we analyze the vertices in the parquet approximation for the weakly interacting Hubbard model
on the square lattice at half-filling, $U/t=2$,
where $t=1$ is the nearest neighbor hopping amplitude.
The temperature is set to $T/t=0.2$. The lattice size is fixed to $8\times8$ sites.

The \textit{victory} implementation of the parquet method 
which we use here was presented in Ref.~\cite{Kauch19}.
It does not make use of the efficient calculation scheme presented in Sec.~\ref{sec:unified},
but it serves us to evaluate the vertices $F$ and $\Phi$ within the parquet approximation.

\begin{figure}
  \includegraphics[width=0.5\textwidth]{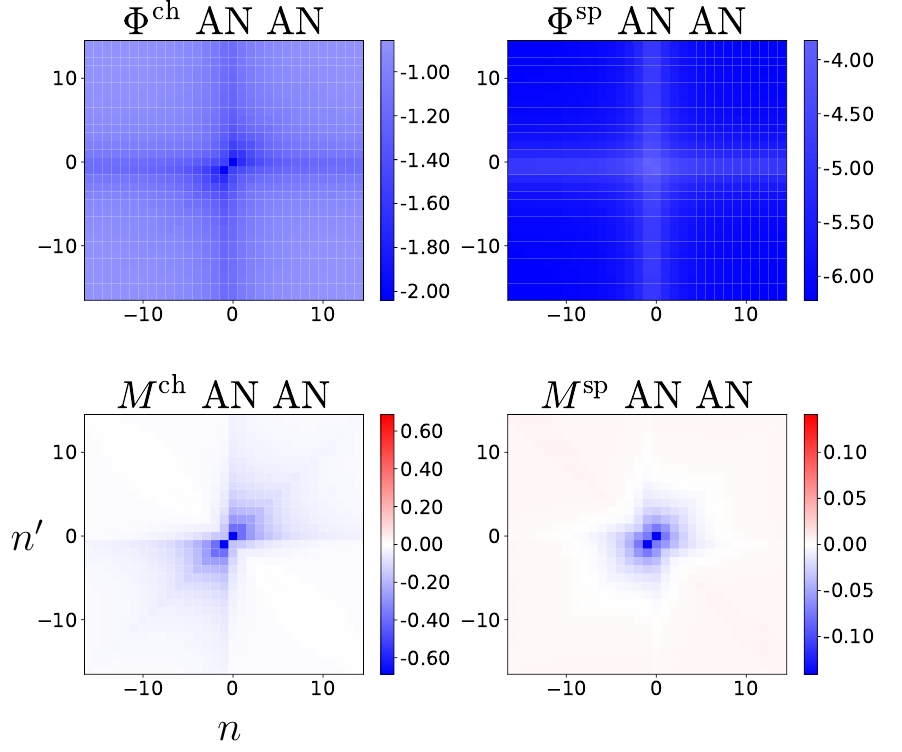}
  \caption{\label{fig:parquet_vertices}
  Reducible vertex functions in the parquet approximation, $U/t=2, T/t=0.2$.
  Fermionic momenta correspond to the antinode (AN),
  {the bosonic momentum is set to} $\qv=(\pi,\pi)$, other labels as in Fig.~\ref{fig:vertices}.
  }
\end{figure}

As mentioned above, in the efficient calculation scheme the parquet
approximation corresponds to setting the fully irreducible vertex
in Eq.~\eqref{eq:uirr_parquet} to zero, $\tilde{\Lambda}^\alpha=0$,
whereas the \textit{victory} implementation actually evaluates Eq.~\eqref{eq:parquet}
using $\Lambda^\alpha=\tilde{\Lambda}^\alpha+U^\alpha=U^\alpha$.
We show in Appendix~\ref{app:m} how the vertices $M$ can be calculated
from the converged solution for $F$ and $\Phi$.
Their full momentum and frequency dependence is available to us,
\begin{align}
M^{\ch/\sz}(\kv,\kv',\qv,\nu,\nu',\omega).
\end{align}

First, we consider the asymptotic behavior of the vertices as a function of the frequencies.
Fig.~\ref{fig:parquet_vertices} shows the particle-hole vertices $\Phi^{\ch/\sz}$ and $M^{\ch/\sz}$,
where we focus on the antinode, $\kv=\kv'=\kv_{\text{AN}}=(\pi,0)$,
the bosonic momentum and frequency are set to $\qv=(\pi,\pi)$ and $\omega=0$.
This combination represents the scattering of particle and hole from the antinode
to another antinode.

Similar to the atomic limit, the $M$'s decay as a function of $\nu, \nu'$
in all directions, but their structure is more complicated due to the additional energy scale $t$.

\begin{figure}
    \includegraphics[width=0.48\textwidth]{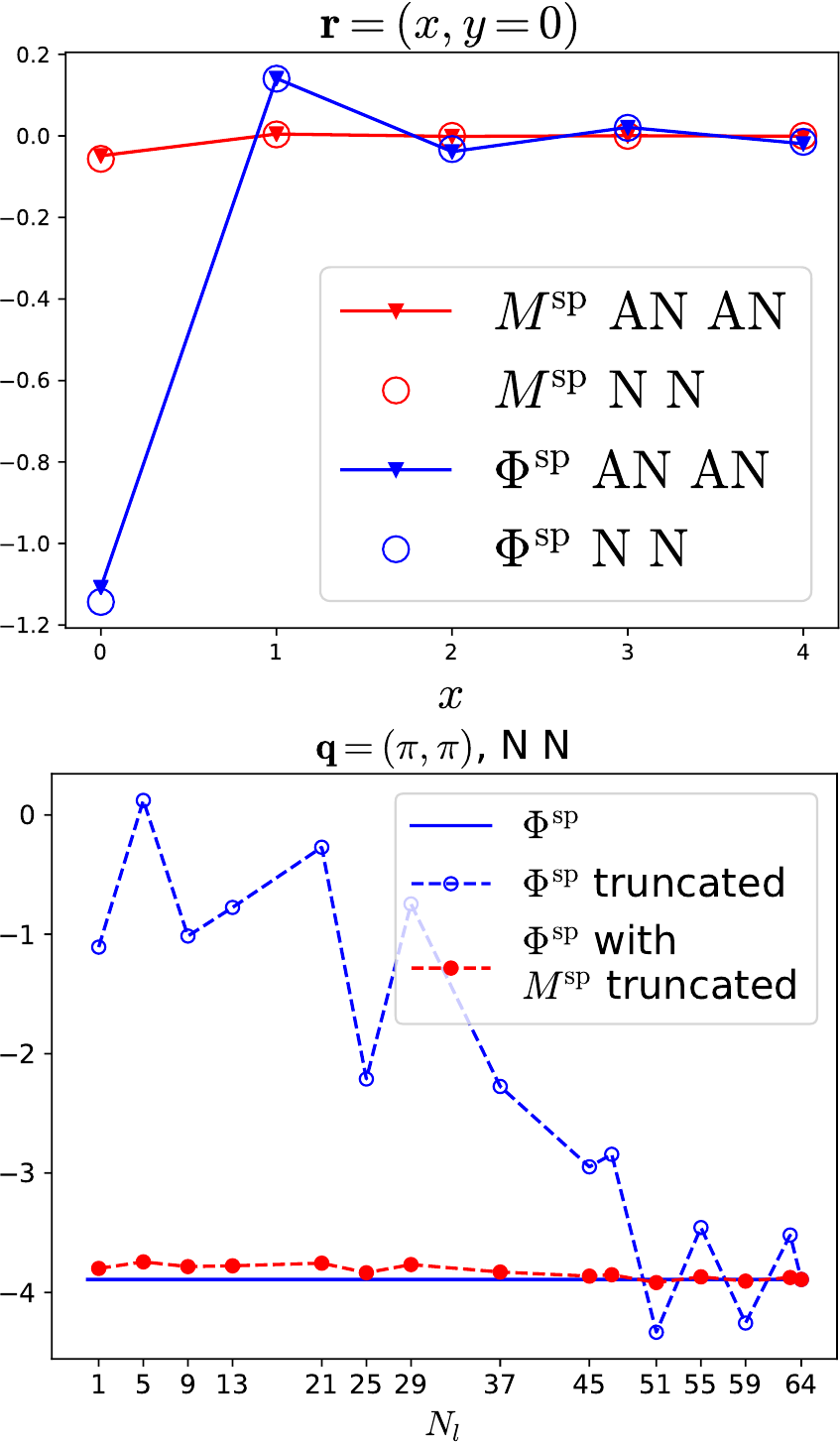}
    \caption{\label{fig:phi_parquet}
    Top: Spatial dependence of $\Phi^\sz$ and $M^\sz$.
    The fermionic momenta correspond to the node or antinode.
    The alternating sign of $\Phi^\sz$ indicates antiferromagnetic correlations.
    {Bottom: Effect of truncation in the form-factor basis on the vertex $\Phi^\sz$.
    Since $M^\sz$ is much smaller than $\Phi^\sz$ the form-factor truncation of $M^\sz$ (red) is quantitatively superior to the direct truncation of $\Phi^\sz$ (blue) [see text].
    Results are for frequencies $\nu=\nu'=\pi T$; $N_\ell=64$ corresponds to a calculation without truncation.
    }
    }
\end{figure}

We note that in the current implementation it is not feasible to fully converge
the Matsubara summations required for the calculation of the vertices $M$
(cf. Appendix~\ref{app:m}).
The correspondence to the calculated $\Phi$ is therefore not perfect
and $M^\sz$ retains a small residual asymptote.

Next, we consider the spatial dependence of the vertices.
To this end, we transform $\Phi^\sz$ and $M^\sz$ to
real space with respect to the bosonic momentum, $\qv\rightarrow\mathbf{r}$,
\begin{align}
M^{\ch/\sz}(\kv,\kv',\mathbf{r},\nu,\nu',\omega).
\end{align}
We fix the frequencies to $\nu=\nu'=\pi T$, $\omega=0$.
For the fermionic momenta we consider the antinode $\kv=\kv'=\kv_{\text{AN}}$
and the node $\kv=\kv'=\kv_{\text{N}}=(\frac{\pi}{2},\frac{\pi}{2})$.

{The top panel of} Fig.~\ref{fig:phi_parquet}
shows $\Phi^\sz$ and $M^\sz$ as a function of
$\mathbf{r}=(x,y=0)$ along the $x$-axis.
Clearly visible is the alternating sign of $\Phi^\sz$
characteristic of antiferromagnetic correlations.
On the other hand, $M^\sz$ is two orders of magnitude smaller than $\Phi^\sz$.
Similarly, in the charge channel $M^\ch$ is much smaller than $\Phi^\ch$ (not shown).
Importantly, this fact alone implies that the spatial dependence of $\Phi=M+\Delta-U$
is largely determined by $\Delta$, the single-boson exchange~\cite{Krien20}.
It explains the fast convergence of the truncated unity approximation~\cite{Eckhardt20}
used in Refs.~\cite{Krien20,Krien20-2}, where only the $M$'s were
truncated in real space while the full spatial dependence of $\Delta$ was retained.

{To underline this,
we transform the vertex $\Phi^\sz$ into the truncated unity (form-factor) basis~\cite{Eckhardt20}
and back into $\qv$-space, while discarding all but a number $N_\ell$ of basis functions (form factors)
$f(\ell,\qv)$,
\begin{align}
\Phi^\sz(\qv,N_\ell)\equiv
\sum_{\ell=1}^{N_\ell}f^*(\ell,\qv)\sum_{\qv'} f(\ell,\qv')\Phi^\sz(\qv'),
\end{align}
where we keep $\nu=\nu'=\pi T, \omega=0, \kv=\kv'=(\frac{\pi}{2},\frac{\pi}{2})$ fixed.

Obviously, $\Phi^\sz(\qv,N_\ell=64)=\Phi^\sz(\qv)$
recovers the complete momentum dependence,
since there are as many form factors as there are lattice sites ($8\times8$).
Blue data points show the result for $\qv=(\pi,\pi)$
in the bottom panel of Fig.~\ref{fig:phi_parquet},
indicating a remarkably slow convergence of the expansion with the cutoff $N_\ell$~\footnote{{The truncated unity approximation applied to fermionic momentum dependence of $\Phi^\mathrm{sp}$ converges much faster in $N_\ell$. This fact was already largely explored in Ref.~\onlinecite{Eckhardt20}. By evaluating the parquet equation [Eq.~\eqref{eq:parquet}] one however effectively approximates also the bosonic momentum (see Ref.~\onlinecite{Eckhardt20} for a detailed discussion).}}.
Apparently, the antiferromagnetic correlations represented by $\Phi^\sz$
should not be truncated in real space even at this high temperature.

{To assess the advantage of the short-range nature of the vertex $M$ in the parquet equation with SBE decomposition [Eqs.~\eqref{eq:ph_kernel_ch}-\eqref{eq:pp_kernel_trip}],
we next apply the same procedure to the vertex $M^\sz$.
The red data points {in the bottom panel of Fig.~\ref{fig:phi_parquet} show the resulting approximation for the thus determined} 
$\Phi^\sz(\qv)\approx M^\sz(\qv,N_\ell)+\Delta^\sz(\qv)-U^\sz$, which is reasonable even for $N_\ell=1$.}

To interpret this result, it is important to remark that
the relative error ${|M^\sz(\qv)-M^\sz(\qv,N_\ell)|}/{|M^\sz(\qv)|}$,
for a given $\qv$, can be similar to ${|\Phi^\sz(\qv)-\Phi^\sz(\qv,N_\ell)|}/{|\Phi^\sz(\qv)|}$.
However, as is clear from the top panel of Fig.~\ref{fig:phi_parquet},
the aim is to capture the coefficients $M^\sz(\ell)=\sum_\qv f(\ell,\qv)M^\sz(\qv)$ which are significant {\sl relative} to $\Phi^\sz$.
Therefore, it is sufficient to keep only a very small number of coefficients $M^\sz(\ell)$, for example, the first form factor, $f(\ell=1,\qv)=1$,
already captures the local component $\sum_\qv M^\sz(\qv)$
drawn in the top panel of Fig.~\ref{fig:phi_parquet} at $\mathbf{r}=(0,0)$.
}

{We should also note that} the $M^\sz$ presented here
was obtained in postprocessing and is not perfectly converged~\footnote{{The original {\it victory} implementation  does not use $W$, $\gamma$ or $M$ vertices as an inherent part of the computation and therefore it is not optimized to obtain them with the same accuracy as $\Phi$. As a consequence, e.g., the tails of $\Phi$ and the tails of $\Delta$ are not treated equally (within the code or in postprocessing, respectively).}}. {Therefore, we can not determine the precise correlation content of this vertex,
for example, whether it is completely free of antiferromagnetic correlations
or instead captures some of them.}
In the future, the implementation of the efficient calculation scheme presented
in Sec.~\ref{sec:unified} may ultimatively clarify this,
because it allows for determining $M$ with the same accuracy as $\Phi$.

{Finally, let us estimate the numerical scaling of the newly proposed scheme as compared to the traditional parquet implementation. For a frequency box of linear size $N_\omega$ and $N_q$ momentum points in the Brillouin zone, the standard parquet calculation requires virtual memory that scales with $\mathcal{O}(N_{q}^3{N}_\omega^3)$ and the computational effort scales with  $\mathcal{O}(N_{q}^4{N}_\omega^4)$ for the Bethe-Salpeter equation~\eqref{eq:bse_t} and with  $\mathcal{O}(N_{q}^3{N}_\omega^3)$ for the parquet equation~\eqref{eq:parquet}. In practice, however, large vertices $\Phi$ need to be stored in distributed memory and internodal communication and memory access operations needed in evaluating Eq.~\eqref{eq:parquet} are the actual bottleneck. }

{In the scheme proposed here the $M$ vertices need only a much smaller frequency box $\widetilde{N}_\omega$. The actual memory requirement $\mathcal{O}(N_{q}^3\widetilde{N}_\omega^3)$ is thus reduced. Additionally, we need to store three-leg vertices $\gamma$ which scale like $\mathcal{O}(N_{q}^2\widetilde{N}_\omega^2)$. Using the form-factor basis to represent the momentum dependence of $M$, this scaling is further reduced to $\mathcal{O}(N_{\ell}^3\widetilde{N}_\omega^3)$ for the $M$'s.
With just few (or even only one as Fig.~\ref{fig:phi_parquet} demonstrates) form factors and small $\widetilde{N}_\omega$, the dominant part is the quadratic scaling in $N_{q}\widetilde{N}_\omega$ for $\gamma$'s. The computational effort of the new Bethe-Salpeter equation [Step 4, Eqs.~\eqref{eq:mbe_ph_2}-\eqref{eq:mbe_pp_2}] scales then with  $\mathcal{O}(N_{\ell}^4\widetilde{N}_\omega^4)$ and the parquet equations in form-factor basis [Step 3, Eqs.~\eqref{eq:ph_kernel_ch}-\eqref{eq:pp_kernel_trip}] with $\mathcal{O}(N_{\ell}^6\widetilde{N}_\omega^3$). Due to significantly smaller vertices $M$ and $\gamma$, the memory access bottleneck can be removed (the $\Delta$'s do not need to be stored)~\footnote{{The advantage of this scheme in comparison with the TUPS implementation of Ref.~\cite{Eckhardt20}, which also uses form-factors, is twofold: (i) we can use significantly fewer form factors and also transform the bosonic momentum;
(ii) the frequency box can be chosen smaller.}}}.

{
\subsection{{Comparison to} $GW$ approximation}
\label{sec:GW}
We now draw a connection between the parquet and $GW$ approximations,
paying special attention to the role of vertex corrections.
To this end, we recall Eq.~\eqref{eq:cc_hedin} for the parquet self-energy,
which is drawn as a diagram at the bottom of Fig.~\ref{fig:polarization}.

Let us examine the effect of dropping the vertex corrections in different places.
The most straightforward way to do this is to set $\gamma=1$ only in the defining equation for $\Sigma$, which is then given as
\begin{align}
    \Sigma^{GW}_k-\Sigma^H=-\frac{1}{2}\sum_q G_{k+q}\left[W^\ch_q+W^\sz_q\right],\label{eq:gw}
\end{align}
where $G$ and $W$ are the Green's function and the screened interaction corresponding to the parquet approximation.
The thus defined $\Sigma^{GW}$ is shown in Fig.~\ref{fig:gw} (cyan)
next to the complete parquet self-energy (red), where $U/t=2, T/t=0.2$, as before. 
Apparently, for these parameters the direct contribution of vertex corrections
to the self-energy is not very large and thus $\Sigma^{GW}$ is still a reasonable approximation.
The inset of Fig.~\ref{fig:gw} shows that
$\gamma^\ch$ and $\gamma^\sz$ deviate from their
noninteracting value $1$ by roughly up to 30\% and 15\%, respectively.
The bottom panel of Fig.~\ref{fig:wgamma} shows that $\gamma^{\ch/\sz}$
are suppressed mainly around the bosonic momentum $\qv=(\pi,\pi)$.

\begin{figure}
    \includegraphics[width=0.5\textwidth]{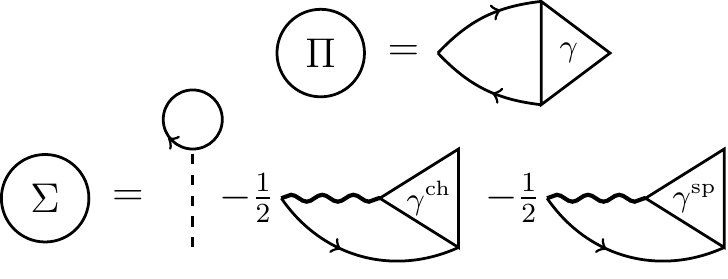}
    \caption{\label{fig:polarization}
    {Hedin's equations for the polarization (top) and the self-energy (bottom).
    Neglecting vertex corrections corresponds to setting $\gamma=1$.}
    }
    \end{figure}

\begin{figure}
  \includegraphics[width=0.48\textwidth]{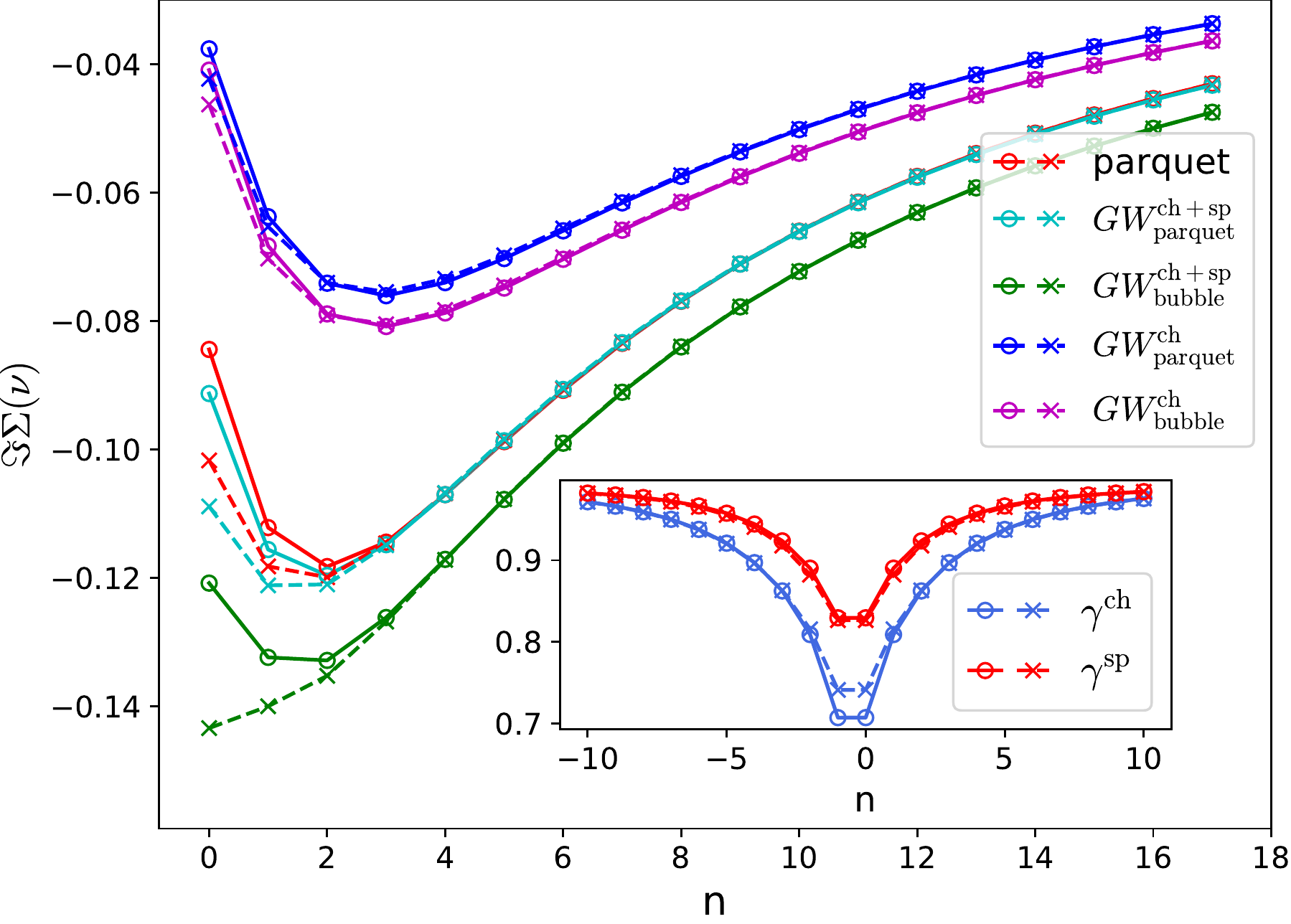}
  \caption{\label{fig:gw}
  {Imaginary part of the self-energy {comparing the parquet approximation (red) with various $GW$-like approximations (see text;  purple:  actual $GW$ approximation)
  for the half-filled square lattice Hubbard model at $U/t=2, T/t=0.2$
  as a function of the Matsubara index.
  Two momenta are shown, corresponding to the node (empty circles) and antinode (crosses).}
  Inset: Hedin vertex $\gamma^\ch$ (blue) and $\gamma^\sz$ (red)
  as a function of the fermionic frequency.
  Fermionic momenta correspond to the node (empty circles) or antinode (crosses) and the bosonic momentum and frequency are set to ${\bf q}=(\pi,\pi)$ and $\omega=0$.}
  }
\end{figure}

\begin{figure}
  \includegraphics[width=0.48\textwidth]{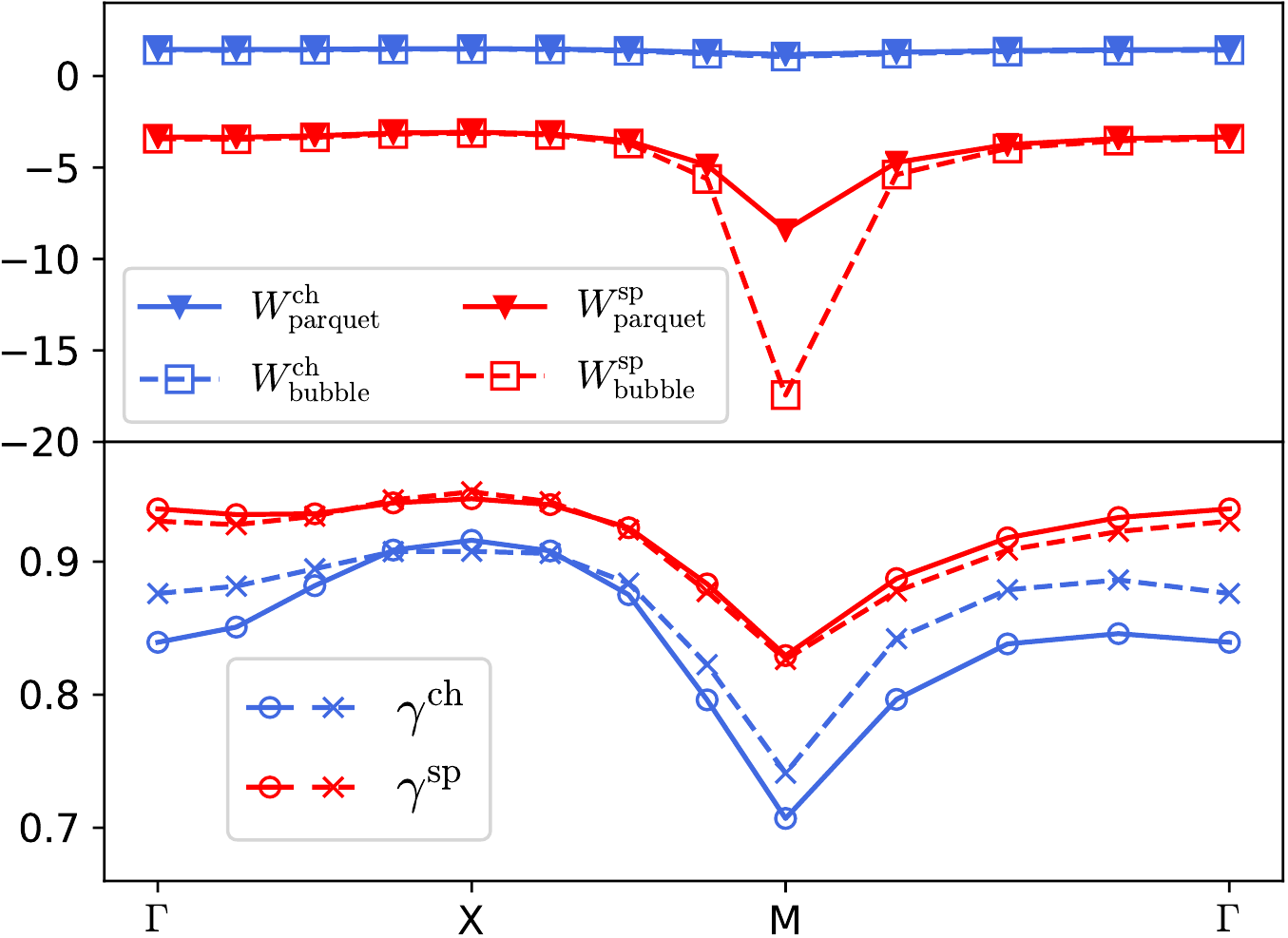}
  \caption{\label{fig:wgamma}
  {Top: Screened interaction in the parquet approximation (full lines)
  and neglecting vertex corrections [cf. Eq.~\eqref{eq:pi_gw}] (dashed lines) as a function of $\qv$.
  Bottom: Hedin vertex $\gamma^\ch$ (blue) and $\gamma^\sz$ (red).
  The fermionic momentum $\kv$ corresponds to the node (empty circles) or antinode (crosses); $\nu=\pi T, \omega=0$. Parameters as in Fig.~\ref{fig:gw}.}
  }
\end{figure}

One needs to keep in mind, however, that the vertex corrections appear
in all diagrammatic objects drawn in Fig.~\ref{fig:polarization} (i.e., $G,W,\gamma$).
Therefore, to approach a $GW$-like approximation of any practical value,
we need to drop further vertex corrections.
For example, let us recall that the screened interaction defined in Eq.~\eqref{eq:cc_dyson_w}
incorporates vertex corrections via the polarization,
which is drawn as a diagram on the top of Fig.~\ref{fig:polarization}.
Consequently, {for an actual $GW$ calculation one should use here} only a bubble of parquet Green's functions for $\Pi$,
\begin{align}
    \Pi^{GW}_q=\sum_k G_k G_{k+q}.\label{eq:pi_gw}
\end{align}
The resulting screened interactions (dashed lines) are drawn in the top panel of Fig.~\ref{fig:wgamma} in comparison to the parquet approximation (full lines).
While $W^\ch$ is similar to the parquet result (but anyways almost momentum independent),
the large difference for $W^\sz$ reveals the significant screening facilitated by $\gamma^\sz$.

{If we use the bubble back in Eq.~\eqref{eq:gw}, the (absolutely) much larger $W^\sz$  in the bubble approximation for the screening leads to a huge feedback on the self-energy:  Fig.~\ref{fig:gw} (green) shows
even an insulating-like behavior} at the antinodal point,
far above the temperature where this is expected to happen~\cite{Schaefer20,Krien20-2}.

The $15\%$-suppression of $\gamma^\sz$ shown in the inset of Fig.~\ref{fig:gw}
therefore crucially determines the Stoner enhancement $(1-U^\sz\Pi^\sz)^{-1}$
in the proximity of the spin-density wave.
This suppression is the result of the particle-particle vertex correction~\footnote{
{See Refs.~\cite{Kitatani19,Krien20} for a calculation of the particle-particle vertex correction in the Anderson impurity model.}}
considered by Kanamori~\cite{Kanamori63}.
To arrive at a reasonable approximation for $W^\sz$ this effect
needs to be taken into account in some way,
for example, by replacing $U^\sz$ with an effective interaction,
which is the essence of the two-particle self-consistent approach~\cite{Vilk97}
and of the Moriya-$\lambda$ correction~\cite{Moriya85,Katanin09}.

Finally, we note that in this work we employed the Fierz ratio $\frac{1}{2}$ for the self-energy
in Eq.~\eqref{eq:cc_hedin}, which corresponds to a symmetric splitting between the charge and
spin channels. This is a natural choice because it leads to a cancellation of
slowly decaying Matsubara summations~\cite{Krien19-3} in Eq.~\eqref{eq:cc_hedin}.

{Usually, however, in $GW$} the self-energy is expressed only through
$W^\ch$ (and $\gamma^\ch$)~\cite{Hedin65}, which, at first glance, 
seems useful to avoid problems due to the instability in the spin channel.
However, Fig.~\ref{fig:gw} shows that the corresponding result for the self-energy
using only $W^\ch$ from the parquet calculation (dark blue) is significantly worse
than the symmetric approximation (cyan) in Eq.~\eqref{eq:gw},
confirming that for an optimal result the channels should be mixed~\cite{Ayral17,Schaefer20-2}.
Also, using the noninteracting Green's function $G^0$
to calculate $\Pi^{GW}$ and $\Sigma^{GW}$ is even worse, as $W^\sz$ is then
already outside of its convergence radius for these parameters (not shown).
We should note that the decoupling ambiguity is a
peculiarity of the Hubbard interaction $Un_\uparrow n_\downarrow$.
It does not affect nonlocal interactions between charge or spin densities.
}

\section{Discussion and Conclusions}\label{sec:conclusions}
The parquet equations for real fermions were reformulated into a computationally more feasible form by combining them with Hedin's $GW\gamma$ formalism.
From the viewpoint of the latter our approach yields the parquet diagrams
for $\gamma$ in terms of single- and multi-boson exchange.
This offers a new perspective on vertex corrections in electronic systems. For example, the association of certain vertex diagrams with effective particles becomes very explicit~\cite{Kauch19-3},
or the notion of a `bosonic glue' that may play a role for phenomena such as
high temperature superconductivity~\cite{Kitatani19} can be taken more literally.

The resulting calculation scheme,
which was coined a boson exchange parquet solver (BEPS) in Ref.~\cite{Krien20},
has no disadvantages in comparison to previous implementations
of the parquet equations but offers two strong advantages.
Namely, the vertex asymptotics and, in the case of a lattice system~\cite{Krien20},
also the long-ranged fluctuations are removed from the parquet
equations through their exact reformulation.

This goes beyond the asymptotic treatment of the vertices pioneered in Refs.~\cite{Kunes11,Li16,Wentzell20}
which improves the feasibility of parquet solvers~\cite{Li16,Tagliavini18,Kaufmannthesis},
but the low-energy aspect of the single-boson exchange
remains intermixed with all other fluctuations.
Instead, the BEPS method corresponds to a kind of separation of the
fluctuations that is exact also at low frequencies~\cite{Krien19,Katanin20}.
In Ref.~\cite{Krien20} and here this idea was adopted to the parquet formalism
for dual fermions and real fermions, respectively.
Let us stress that BEPS for real fermions is an exact unification of Hedin's equations and the parquet equations. An approximation only enters when the fully irreducible vertex $\Lambda$
is replaced by an approximated one such as the bare interaction $U$ in the parquet approximation or all local diagrams in the D$\Gamma$A,~\cite{Toschi07,Valli15,Li16,Kauch19,Ayral16}~\footnote{
{This statement holds notwithstanding cutoff and truncation errors, e.g.,
of Matsubara summations, in a particular implementation.}
}.
{In this respect our approach does not differ from the traditional
parquet method, that is, it does not introduce any additional approximations.}

{As numerical results} we first discussed the simple case of a quantum impurity model,
where the spatial degrees of freedom do not play a role.
The computational efficiency of the calculation scheme is then
improved {through} the decay properties of the vertices.
Remarkably, this is sufficient to solve
the parquet equations for the atomic limit on a laptop
using the provided \texttt{Python} script~\cite{KrienBEPSpython}.

We also analyzed the parquet approximation for the lattice Hubbard model
using the \textit{victory} code presented in Ref.~\cite{Kauch19}.
We evaluated the vertices that correspond to the BEPS method and verified
that they indeed exhibit the useful decay properties.
This underlines the accuracy of the asymptotic treatment
of the vertices in this implementation~\cite{Li16}.
{We discussed $GW$-like approximations for the self-energy,
highlighting the crucial importance of vertex corrections represented
by the Hedin vertex $\gamma$, which even increases at low temperature~\cite{Krien20-2}.
However, we find it plausible that neglecting vertex corrections
has a less severe effect away from particle-hole symmetry and in dimensions $>\!2$.
{With respect to recent works investigating} the feedback of spin fluctuations
on the optical conductivity~\cite{Kauch19-3,Worm20,Simard20},
it is intriguing to consider the role of the
fermion-boson coupling also in this context.
}

In the future we will implement the efficient calculation
scheme into the \textit{victory} code~\cite{Kauch19}.
This seems promising because in Refs.~\cite{Krien20,Krien20-2},
which discussed the lattice case for dual fermions,
it is shown that the BEPS method unfolds its full power in combination with the truncated unity (TU) approximation~\cite{Husemann09,Wang12,Thomale13,Lichtenstein17,Eckhardt20},
which corresponds to a real space truncation of the vertices.
We expect (cf. Fig.~\ref{fig:phi_parquet}) also for real fermions a similar improved convergence with the form factors
compared to the truncated unity parquet solver (TUPS,~\cite{Eckhardt20}).
While the form factors correspond to a suitable basis for the spatial degrees of freedom,
the computational efficiency may be further improved by introducing an optimal basis for the frequencies~\cite{Shinaoka18,Witt20,Wallerberger20}.
{Such a treatment should pave the way for parquet calculations including at least 
a few orbitals  such as the two $e_g$- or three $t_{2g}$-orbitals of
a transition metal oxide, or the $J=5/2$ multiplet of an $f$-electron system.} 

Lastly, some words are in place regarding the closely related method for dual fermions presented in Ref.~\cite{Krien20}: Diagrammatically, both approaches are the same, but the basic building blocks are different. In Ref.~\cite{Krien20} the (real fermion) Green's function lines are replaced by dual fermion lines and the equations defining the self-energy, polarization, and Hedin vertex assume a different form.
In the present paper, the starting point is an approximation for the two-particle fully irreducible vertex $\Lambda$. In contrast, in the dual fermion formulation~\cite{Krien20} local {\sl reducible} interactions are included, hence using, e.g., a local full vertex $F_\text{loc}$ for dual fermions instead of a local $\Lambda_{\text{loc}}$ in D$\Gamma$A.
This leads to an interesting distinction between the bosonization of the
parquet equations for real and dual fermions, respectively,
which corresponds to removing the interaction-reducible diagrams from
either $F_{\text{loc}}$ or $\Lambda_{\text{loc}}$:
In {the} case of real fermions $\Lambda_{\text{loc}}$ includes only one such diagram,
the bare interaction itself,
whereas for dual fermions $F_{\text{loc}}$ contains many interaction-reducible diagrams 
which need to be separated off using the (local) SBE decomposition~\cite{Krien19-2}.

The approaches for real and dual fermions both have their pros and cons. The parquet solver
discussed in the present paper is simpler, an integral part of many different approaches,
the interpretation in terms of real fermions is easier,
and the approximation made for  $\Lambda$ is very explicit.
On the other hand, the connecting dual fermion lines decay much faster,
which, in combination with the decay of the vertex functions facilitated by the BEPS method,
leads to a very high computational efficiency of the dual parquet solver~\cite{Krien20}.
Further, the dual fermions are not affected by divergences of the vertex $\Lambda_\text{loc}$~\cite{Schaefer13}.
{It is noteworthy that, due to the dependence of the bare dual fermion interaction ($F_{\text{loc}}$) on three frequencies,
the (parquet) dual self-energy can not be expressed in terms of
$G, W$, and $\gamma$ alone~\cite{Krien20-2}.
This is possible for the real fermion representation, which allowed us here to to establish a connection between the parquet and $GW\gamma$ methods.
}

\acknowledgments
We thank C. Eckhardt, A. Valli, and M. Wallerberger for useful comments on the text and
S. Andergassen, M. Capone, P. Chalupa,
C. Hille, M. Kitatani, A.I. Lichtenstein, E.G.C.P. van Loon,
G. Rohringer, and A. Toschi for discussions.
The present research was supported by the Austrian Science Fund (FWF)
through projects P32044 and P30997.

\appendix

\section{Evaluation of vertices $M$}\label{app:m}
Here, we show how the vertices $M^{\alpha}$ in the parquet expression~\eqref{eq:uirr_parquet}
can be calculated from a converged result of the \textit{victory} code~\cite{Kauch19},
that is, the Green's function $G$ and the vertices $\Phi$ and $F$
in the parquet decomposition~\eqref{eq:parquet} are known.
We focus on the particle-hole channels $\alpha=\ch,\sz$.
First, we determine the susceptibility and the screened interaction,
\begin{align}
X^\alpha_q=&2\sum_k G_k G_{k+q}+2\sum_{kk'q} G_k G_{k+q}F^\alpha_{kk'q}G_{k'}G_{k'+q},\notag\\
&W^\alpha_q=U^\alpha\left(1+\frac{1}{2}X^\alpha_qU^\alpha\right).\label{app:w}
\end{align}
Next, we evaluate the Hedin vertex $\gamma$.
We insert Eq.~\eqref{eq:nabla} into Eq.~\eqref{eq:gamma_chsp},

\begin{align}
\gamma^{\alpha}_{kq}=&1+\sum_{k'}
(F^{\alpha}_{kk'q}-\gamma^\alpha_{kq}W^\alpha_q\gamma^\alpha_{k'q})G_{k'}G_{k'+q}\notag\\
=&1+\sum_{k'}F^{\alpha}_{kk'q}G_{k'}G_{k'+q}-\gamma^\alpha_{kq}W^\alpha_q\Pi^\alpha_{q}\notag\\
=&1+\sum_{k'}F^{\alpha}_{kk'q}G_{k'}G_{k'+q}-\gamma^\alpha_{kq}\frac{1}{2}U^\alpha X^\alpha_{q}
\label{eq:gamma_red}.
\end{align}

From the first to the second line we identified the polarization $\Pi$ using Eq.~\eqref{eq:pi}.
From the second to the third line we used Eq.~\eqref{eq:w} and $X^\alpha_q=2\Pi^\alpha_q/(1-U^\alpha\Pi^\alpha_q)$.
We solve Eq.~\eqref{eq:gamma_red} for $\gamma$,

\begin{align}
\gamma^{\alpha}_{kq}=&\frac{1+\sum_{k'}F^{\alpha}_{kk'q}G_{k'}G_{k'+q}}{W^\alpha_q/U^\alpha},
\end{align}

where we used again Eq.~\eqref{app:w}.
With $\gamma$ and $W$ we finally obtain $M$ from Eq.~\eqref{eq:phi_m} (see also Fig.~\ref{fig:phi}),
\begin{align}
{M}^{\alpha}_{kk'q}=\Phi^{\alpha}_{kk'q}-\gamma^\alpha_{kq}W^\alpha_q\gamma^\alpha_{k'q}+U^\alpha.
\label{app:phi_m}
\end{align}
We do not evaluate $M^{\sing/\trip}$ in the particle-particle channel. However,
for the singlet channel $\alpha=\sing$ the steps are analogous,
starting from Eq.~\eqref{eq:gamma_singlet} and taking into account the factor $\frac{1}{2}$.
For the triplet channel nothing needs to be done since $M^\trip=\Phi^\trip$.

\bibliography{main}

\end{document}